\magnification=1200 
\baselineskip=3 ex 
\hsize=16 true cm \vsize=22 true cm


\def\ast#1{{ 1\over \pi i}\int_{-\infty}^\infty\,{ d#1
\over #1 -i}\,}

\def\c#1{{\cal#1}}
\def\cc#1{\v C^#1}

\def\cl{\nabla \times}
\def\d3#1{d^3\v#1\,}
\def\dd{\,\partial{\kern-1.3ex^{^\leftrightarrow}}\!\!}

\def\dir{\partial\kern -1.2ex /\,}
\def\dpt{d\t p\,}
\def\dv{\nabla\cdot }

\def \gr{\nabla}

\def\h#1{\hat#1}

\def\ip#1#2{\v #1\cdot\v #2}
\def\lt{\langle\, }
\def\monthname{\ifcase\month\or January\or February\or
March\or April\or May\or June\or July\or August\or
September\or October\or November\or December\fi}
  \def\p{\partial}

\def\rr#1{{\v R^#1}}
\def\rt{\,\rangle}
\def\skp{\vskip 2.5ex}

\def\t#1{\tilde#1}

\def\tp{{(2\pi )}} 
\def\v#1{{\bf#1}}



\centerline{\bf Wavelet Electrodynamics I}
\centerline{(Appeared in Physics Letters A {\bf 168} (1992) 28-34.)}
\skp
\centerline {\bf  Gerald Kaiser } 
\centerline {Department of Mathematics} 
\centerline {University of Massachusetts at Lowell} 
\centerline {Lowell, MA 01854, USA} 
\skp
\skp

\centerline {\bf ABSTRACT}

\skp
\noindent 
A new representation for solutions  of Maxwell's equations 
 is derived.  Instead of being expanded in  plane  waves, the solutions are
given as linear  superpositions of {\sl spherical  wavelets\/} dynamically
adapted to the Maxwell field and well--localized  in space at the initial
time.   The wavelet  representation  of a  solution is analogous to its Fourier
representation,   but has the  advantage of being {\sl local.\/}  It is closely
related to the relativistic coherent--state representations for the
Klein--Gordon and Dirac fields developed in earlier work.

\skp

\noindent  Key words: Electrodynamics, Maxwell's equations, wavelets,
coherent states.

\skp\skp

\centerline{ \S 1.  \bf Introduction} 

\skp

\noindent In previous work [1,2] coherent--state representations have
been  developed for various relativistic systems, including Klein--Gordon
and Dirac fields and their wave functions. The main tool was the {\sl
analytic--signal transform\/}  (AST) [3,4], which gives a canonical extension
of fields from real spacetime to complex spacetime.  When the field $f(x)$ is
{\sl free,\/}  its extension $f(z)$ is analytic in the {\sl double tube\/}
domain

$$
 \c T=\{z\equiv (\v z, z_0)=x-iy\in\cc4\,|\,y^2\equiv y_0^2-\v y^2>0\}=\c
T_+\cup\c T_-,
\eqno(1)
  $$
where $\pm y_0>0$ in $\c T_\pm$.  The restriction of $f(z)$ to the forward
tube $\c T_+$ ($y_0>0$) then contains only the positive--frequency part of
$f$, while its restriction to the backward tube $\c T_-$ ($y_0<0$) contains
only the negative--frequency part. The points of $\c T$ parametrize a
system of relativistic coherent states $e_z$, with $z\in\c T_+$ and $z\in\c
T_-$ representing particles and antiparticles, respectively, and it was shown
that the state $e_z$ labeled by $z=x-iy$ has an expected position $\v x$ at
time $x_0$ and an expected energy--momentum  proportional to $y$. Certain
six--dimenional submanifolds of $\c T$ can therefore be interpreted as {\sl
phase spaces.\/}   Furthermore, when the  field has a positive mass,  it can
be reconstructed from its values on any one of these phase spaces.  However,
this reconstruction fails when the mass vanishes, due to the divergence of
the relevant integrals.  (The massless representation is not
square--integrable with respect to the `Liouville measure' on phase space.) 
In this paper we develop a square--integrable representation for the most
common massless system, namely the electromagnetic field.  This will be
done by reducing the dimenionality of the ``phase space'' from six to {\sl
four.\/}  Namely, an electromagnetic field will be reconstructed from its
values in the {\sl Euclidean region\/} (real space--  and imaginary
time--coordinates), as obtained by applying the AST to the field in real
spacetime.  The result will be seen to be a multidimensional generalization
of {\sl wavelet analysis.\/}  

 That relativistic coherent states behave like wavelets due to
Lorentz contractions has been noted in [3].  However, only in the
massless case can the correspondence be complete, since a positive mass
provides a scale, namely the width of $e_z$ in its rest frame. This
led us to the expectation that massless fields (where no rest frames and
canonical scales occur) possess a natural wavelet representation [4].  The
simplest case is the  wave equation in two spacetime dimensions,  for which
a wavelet representation was indeed constructed in [5].  As shown there, the
wavelets are covariant not only under the group of affine transformations
(translations and dilations) but also under the larger group of conformal
transformations.  Here we develop the simplest aspects of a similar
construction for the electromagnetic field in four spacetime dimensions.  A
more detailed analysis of this and related topics, such as conformal
invariance and the construction of {\sl polarized\/} electromagnetic
wavelets,  will be dealt with in a forthcoming paper [6].  To the author's
knowledge, the results given in [5] and in the present paper represent the
first successful ``relativistic coherent--state'' formulations of massless fields
(in two and more spacetime dimensions, respectively).

It is remarkable that a single function, namely the AST--extended field
$f(z)$, combines the  concept of {\sl phase space\/}  (parametrized
by position and momentum) with the concept of {\sl wavelet space\/} 
(parametrized by position and scale).   This unification is actually the 
``relativistic dividend'' earned from the construction of relativistic coherent
states, in the same way as relativity unifies energy with momentum and
electric fields with magnetic fields.  To see this, note that although $y$ is
proportional to the expected energy--momentum of $e_z$, it cannot {\sl
be\/} an energy--momentum since it has units of length (which scale
oppositely to those of momentum).  Rather, $\v y/y_0$ gives the expected
velocity, while the imaginary time $y_0$ gives a {\sl scale parameter.\/}   In
the Euclidean region, $\v y=\v 0$, which means that $e_z$ represents a {\sl
spherical wave\/} which first implodes towards a point in space, then
explodes away from it.  Then $y_0$ measures the scale of this wave by
giving its diameter at the instant of maximal localization.
 Such $e_z$'s will form our system of electromagnetic wavelets in the next
section, and the field itself will be constructed from them.  The relation of
our formalism to the usual one--dimensional wavelet analysis is discussed in
Section 3.

We consider solutions  of Maxwell's
equations in free spacetime $\v R^4$.  It will be convenient to unify the
electric and magnetic fields $\v E(x)$ and $\v B(x)$ into a single {\sl
complex\/}  vector field $\v F(\v x, t)\equiv \v F(x)=\v E(x)+i\v
B(x)$, which then satisfies (with the speed of light  $c=1$)

$$
  \dv \v F=0,\quad i\p_t\v F=\cl \v F.
\eqno(2)
  $$
Let us review the usual solution using  Fourier transforms.  The
above equations imply that $\v F$ satisfies the wave equation
$\p_t\,^2\v F=\gr^2\v F$, hence it has the form

$$\eqalign{
 \v F(x)=\int_C \dpt e^{-ipx }\,\v f(p), 
\cr}
 \eqno(3)
  $$
where $p^2=p_0^2-\v p^2, \  px=p_0 t-\v p\cdot \v x,$ 
$C=C_+\cup C_-$ is the double light cone with $C_\pm=\{(\v p,
p_0)\,|\,\pm p_0=|\v p| >0 \}$  (the origin $p=0$ is excluded) and
$\dpt\equiv \tp^{-3}\d3p/2|\v p| $ is the Lorentz-invariant measure
on $C$.  Maxwell's equations imply  $\v p\cdot \v f(p)=0$ and $ p_0\v
f(p)=i\v p\times\v f(p)$  for $p$ in $C$, which can be solved by
introducing a four--potential.   A Poincar\'e--invariant norm (and associated
inner product) on solutions is given [7] in momentum space by

$$
  \|\v f\|^2\equiv \int_C  \dpt \omega ^{-2} |\v f(p)|^2,
\eqno(4)
  $$
where $\omega \equiv |\v p|$, and we denote the Hilbert space
of all solutions with finite norm by $\c H$.  Gross [7] has shown that a
 norm unitarily equivalent to the above is also invariant under the
fifteen--dimensional {\sl conformal group\/}  $\c C$ of spacetime, which is
generated by the Poincar\'e group together with uniform dilations
($x\to\alpha x,\, \alpha >0$) and inversions in the unit hyperboloid ($x\to
x/x^2$).  This gives a unitary representation  of  $\c C$ on $\c H$.  (That
Maxwell's equations are invariant under $\c C$ has been known for a long
time [8]; the unitary representation gives that invariance a
quantum--mechanical flavor, since the elements of $\c H$ may now be
regarded as single--photon wave functions.)  The new norm has the
following non--local expression  in terms of the fields at time $t=0$:

$$
 \|\v F\|^2={1 \over \pi ^2}\int_{\v R^6}{\d3x \d3y 
\over |\v x-\v y|^2}\,\v F(\v x, 0)^*\cdot \v F(\v y, 0),
\eqno(5)
 $$
where the asterisk denotes complex conjugation.  In the next section we
shall find a simpler expression for $\|\v F\|^2$ in terms of the values $\v
F(\v x, -is)$ of the field in  {\sl Euclidean spacetime\/} (real space but
imaginary time).  This will lead us directly to the wavelet expansion  of $\v
F$.

\skp\skp

\centerline{ \S 2.  \bf The Wavelet Representation of Solutions} 

\skp

\noindent Our extension of $\v F (\v x,t) $ to  complex spacetime employs
the {\sl analytic--signal transform,\/}  developed in [1,2,3] and
further applied in [4,5].  Given an arbitrary but reasonable function $f(\v
x)$ on $\v R^n$ (a smooth function with mild decay is more than sufficiently
reasonable), its analytic--signal transform is the function $f(\v x+i\v y)$ on
$\v C^n$ defined by

$$
 f(\v x+i\v y)=\ast \tau f(\v x+\tau \v y).
\eqno(6)
 $$
In terms of the Fourier transform $\h f$ of $f$,

$$
 f(\v x+i\v y)=(2\pi )^{-n} \int d^n\v p\,2\theta (\ip py)\,e^{i\v
p\cdot(\v x+i\v y)}\,\h f(\v p),
  \eqno(7)
 $$
where $\theta $ is the step function ($\theta (u)=1 $ if $u>0$, $\theta
(0)=1/2$ and $\theta (u)=0$ if $u<0$). Although $f(\v x+i\v y)$ is in general
not analytic, it can be easily shown to be {\sl partially\/}  analytic in the
direction of $\v y\ne\v 0$.  Furthermore,  if $\h f(\v p)$ has certain
support properties (vanishes outside of a solid double cone $V\subset \rr
n$), then  $f(\v x+i\v y)$ is analytic in a corresponding domain in $\cc n$
(the double tube  over the double cone $V'$ dual to $V$).  This will
be seen explicitly below, where $\rr n$ is spacetime.

  Application of the AST to the electromagnetic field $\v F(x)$ gives

$$\eqalign{
\v F(x-iy)\equiv \ast \tau  \v F(x-\tau y)
=\int _C \dpt 2\theta (py)\,e^{ -ip(x-iy)} \,\v f(p).
 \cr}
 \eqno(8)
 $$
(We write $x-iy$ instead of $x+iy$ to conform with the convention used in
Streater and Wightman [9].)  If $y$ belongs to the  forward  solid light cone
$V'_+$ (i.e., $y^2>0$ and $y_0>0$), then $\theta (py)\equiv 0$ on $C_-$ and
$\theta (py)\equiv 1$ on $C_+$, hence $\v F(x-iy)$ is analytic at $x-iy$ and
contains only the positive--frequency part of the field.  Similarly, if $y$
belongs to the backward solid light cone $V'_-$, then $\v F(x-iy)$ is analytic
at $x-iy$ and contains only the negative--frequency part of the field.  Thus
$\v F(x-iy)$ is analytic in the  double tube $\c T$ defined in Eq. (1).
The jump discontinuity in $\v F(x-iy)$ between the future and past
imaginary directions is related to the multidimensional Hilbert transform of
$\v F(x)$  (see [4,5] for details).   In this paper, we shall be
interested only in the transform of $\v F(\v x, t)$ to complex time $t\to
t-is$ (i.e., $\v F(x-iy)$ with $y=(\v 0, s)$):

$$\eqalign{
\v F(\v x, \, & t-is)=\int_C \dpt 2\theta (p_0s)\,e^{ -ip_0(t-is)+
 i\ip px} \,\v f(p)\cr
&=\int_\rr 3 { \d3p\over \tp^3\omega }\, e^{i\ip px}
\,\left[\theta (s) \,e^{-\omega (s+it)}\,\v f(\v p,\omega ) 
+\theta (-s)\,e^{\omega (s+it)}\,\v f(\v p, -\omega )\right]\cr
&=\left\{\omega^{-1}\left[\theta (s) \,e^{-\omega (s+it)}\,\v f(\v p,\omega ) 
+\theta (-s)\,e^{\omega (s+it)}\, \v f(\v p, -\omega )\right] \right\}
\check{\ }(\v x),
 \cr}
 \eqno(9)
  $$
where $\check{\ }$ denotes the inverse Fourier transform in $\rr3$.  Note
that $\v F(\v x, t-is)$ is analytic in $t-is$ whenever $s\ne 0$.    The last line
in Eq. (9) implies, by  Plancherel's theorem, that for $s\ne 0$,

$$\eqalign{
 \int \d3x & |\v F(\v x, t-is)|^2\cr
&=\int_\rr 3  {{ \d3p\over \tp^3 \omega ^2}}
\left|\theta (s) \,e^{-\omega (s+it)}\,\v f(\v p,\omega ) 
+\theta (-s)\,e^{\omega (s+it)}\, \v f(\v p, -\omega )\right| ^2\cr
&=\int_\rr3{\d3p \over \tp^3\omega ^2}
\left\{\theta (s)\,e^{-2\omega s}\,|\v f(\v p, \omega )|^2+
 \theta (-s)\,e^{2\omega s}\,|\v f(\v p, -\omega )|^2 \right\},
 \cr}
 \eqno(10)
  $$
and therefore

$$\eqalign{
\int_\rr 4  \d3x ds\,|\v F(\v x, t-is)|^2&=
\int_\rr3{\d3p \over \tp^3  2\omega ^3} \left\{|\v f(\v p, \omega )|^2+
  |\v f(\v p, -\omega )|^2\right\}\cr
&=\int_C \dpt \, \omega  ^{-2}\,|\v f(p)|^2=\|\v f\|^2.
 \cr}
 \eqno(11)
 $$
Hence we define

$$
 \|\v F\|^2\equiv \int_\rr 4 \d3x ds\,|\v F(\v x, t-is)|^2,
\eqno(12)
  $$
so that $\|\v F\|=\|\v f\|$.  Note that the new norm $\|\v F\|$   and its
associated inner product $\lt\v G\,|\,\v F\rt$ are {\sl local\/}  in the
Euclidean spacetime variables $(\v x, \, s)$.    Define the function $\h e_{\v
y, t-is} (p)$ on $C$ by its complex--conjugate as

$$
\h e_{\v y,t-is}(\v p,p_0)^*=\omega ^2\,2\theta(p_0s)\,e^{-ip_0(t-is)+i\ip py }
.  \eqno(13)
  $$
 The spacetime function corresponding to $\h e_{\v y, - is} $ is

$$\eqalign{
 e_{\v y, - is}(\v x, t) &\equiv \int_C \dpt e^{-ipx} \,
\h e_{\v y, -is} (p)\cr
&=\int_C\dpt \omega^2\,2\theta(p_0s)\,e^{-p_0(s+it)+i\v p\cdot(\v x-\v
y)}.
 \cr}
 \eqno(14)
 $$
This is a scalar solution of the {\sl wave equation\/} which depends only
on $|\v x-\v y|$ and $t-is$, being analytic in the latter variable whenever
$s\ne 0$.  Although $e_{\v y, t- is} $ does not belong to $\c H$ (since it 
 has no polarization, being a scalar),  we shall write Eq. (9) as

$$\eqalign{
 \v F(\v x, t-is)=\lt \h e_{\v x, t-is} \,|\,\v f\rt
=\lt e_{\v x, t- is} \,|\,\v F\rt,
 \cr}
 \eqno(15)
 $$
where $\lt\,|\,\rt$ denotes the inner product in $\c H$, expressed either in
momentum space or, equivalently, in Euclidean spacetime.    Then it follows
from $\|\v F\|^2=\|\v f\|^2$  that the inner product of two solutions $\v G$
and $\v F$  in $\c H$ can be written as

$$\eqalign{
\lt\v G\,|\,\v F\rt &\equiv \int_\rr4 \d3y ds\,\v G(\v y,
is)^*\cdot \v F(\v y, is)\cr
 &=\int_\rr4 \d3y ds\, \lt\v G\,|\,e_{\v y,-is} \rt\lt e_{\v y, -is} \,|\,\v F\rt,
 \cr}
 \eqno(16)
 $$
which gives the {\sl continuous resolution of unity\/} 

$$
 \int_\rr4 \d3y ds\,\,|\,e_{\v y, -is} \rt\lt e_{\v y, -is} \,|\,=I_{\c H},
\eqno(17)
  $$
where $I_{\c H}$ is the identity operator in the space of solutions and the
equality holds in a {\sl weak\/}  sense.  Hence for $\v F$ in $\c H$  we have

$$\eqalign{
\v F(\v x, t-i\sigma )=\lt e_{\v x, t-i\sigma } \,|\,\v F\rt
=\int_\rr4 \d3y ds\,\lt e_{\v x, t-i\sigma } \,|\,e_{\v y, -is} \rt \lt e_{\v y,
-is} \,|\,\v F\rt.
 \cr}
 \eqno(18)
 $$
  A straightforward computation gives

$$
 \lt e_{\v x,t-i\sigma}\,|\,e_{\v y, -is}\rt=
{2\theta (\sigma s) \over \pi ^2}\,{ 3\tau ^2-r^2\over (\tau ^2+r^2)^3},
\quad \tau \equiv s+ \sigma +it,\ r\equiv |\v x-\v y|.
\eqno(19)
  $$
 To obtain the wavelet expansion of  $\v F(\v x, t)$,  we simply take
$\sigma =0$ in Eq. (18):

$$\eqalign{
 \v F(\v x, t)&=\int_\rr4 \d3y ds\,  \lt e_{\v x, t}\,|\,e_{\v y, -is} \rt \,\v
F(\v y, -is) \cr
&=\int_\rr4 \d3y ds\,\, e_{\v y, -is}(\v x, t)  \,\v F(\v y, -is).
 \cr}
 \eqno(20)
 $$
 As we have seen, $e_{\v y, -is} (\v x, t)$ satisfies the
wave equation in $(\v x, t)$.  Since it depends only on $r=|\v x-\v y|$, it is
a {\sl spherical wave\/}  centered at $\v x=\v y$.  Furthermore, $s$ acts as
a {\sl scale parameter,\/}  since 

$$
 e_{\v y, -is} (\v x, t) 
=s^{-4}\,e_{\v y/s, -i} ( \v x/s,  t/s) .
\eqno(21)
  $$
Hence it suffices to examine any one of the wavelets $ e_{\v y, -is}$.  Figures
1-4 show  the behavior of the wavelet centered at the origin with scale
$s=-1$, i.e. of the {\sl basic wavelet\/} 

$$
 w(r, t)\equiv \pi ^2e_{\v 0, i} (\v x, t)={3(1-it)^2-r^2 \over [(1-it)^2+r^2]^3},
\quad r\equiv |\v x|.
\eqno(22)
  $$
These figures confirm that for $t<0$, $e_{\v y, -is}(\v x, t)$  is an incoming
spherical wave which builds up rapidly (at the speed of light!) to a
well--localized packet in the   ball  $|\v x-\v y|\le \sqrt{3}\,|s|$ and decays
rapidly for $t>0$ into an outgoing spherical wave.  The characteristics
$t=\pm r$ appear as ripples in figures 1-3.  These properties partly justify
our use of the term ``wavelets.''  Further justification is given in the next
section.

\skp
\skp

\centerline{\bf 3.   Relation to Wavelet Analysis} 

\skp

\noindent  The correspondence $\v F (\v x, t) \to\v F(\v y, is)$ is analogous
to the Fourier transform $\v F (\v x, t)  \to\v f(p)$, and the
reconstruction of $\v F(\v x, t) $ from $\v F(\v y, is)$ is analogous to its
reconstruction from $\v f(p)$ via the inverse Fourier transform.  The role
of the plane waves $e^{ipx }$ is now played by  $e_{\v y, -is}$, and the local
nature of these functions means that the behavior of  $\v F(\v y, is)$  is
 correlated with that of the  field $\v F (\v x, t)$ in real spacetime,
the correlation being strongest when $t=0$ since  $e_{\v y, -is}$ is then most
localized.  [By contrast, the Fourier transform has no such local property
since the plane waves  extend to all of spacetime;  thus a small local
perturbation in $\v F (\v x, 0)$  can cause an unrecognizable change in $\v
f(p)$.]  In fact, $\v F(\v y, is)$ is a  version of  $\v F(\v x, 0)$, {\sl
blurred to resolution $|s|$.\/}  Eq. (20) with $t=0$ states that $\v F(\v x, 0)$
is recovered by superposing wavelets with different centers $\v y$ and
scales $s$,  with $\v F(\v y, is)$ as the coefficient function. This
representation of functions as superpositions of their blurred versions  is
typical of  wavelet analysis  [10,11,12] and the related 
multiresolution analysis  [13,14].  Moreover, in our case the wavelet
representation  also gives the time evolution because our
wavelets are {\sl dedicated\/} [4,5] to the dynamics of the electromagnetic
field.

 In momentum space, the wavelet with $s=1$ and $\v
y=\v 0$ is represented by

$$
 \h e_{\v 0, -i}(\v p, p_0)=2\omega ^2\theta (p_0)\,e^{-p_0},\quad  p\in C.
\eqno(23)
  $$
This is a multi--dimensional generalization of a basic wavelet which has
 appeared previously in the literature in connection with the
usual (one--dimensional) wavelet theory [15,16], and also in a wavelet
analysis of solutions of the wave equation in two spacetime dimensions (Ref.
[5], Eq. 161).  In fact, the AST is a special case of a {\sl windowed Radon
transform,\/}  which in turn has been shown to be a multivariate
generalization of the wavelet transform.

\skp\skp

\centerline{\bf Acknowledgements} 

\skp
\noindent
I thank my colleagues Ron Brent and James Graham--Eagle for their help
with the  graphics, and the referree for remarks which helped to clarify the
exposition.

\vfill\eject
\centerline{\bf References} 

\skp

\item{[1]} G. Kaiser,  Phase--Space Approach to Relativistic Quantum 
Mechanics.  Part I: Coherent--State Representation  of the Poincar\'e
Group,  J. Math. Phys. {\bf 18} (1977), 952-959;
part II: Geometrical Aspects, {\sl ibid.\/}  {\bf 19} (1978), 502-507;
part III: Quantization, Relativity, Localization and Gauge Freedom,  {\sl
ibid.\/} {\bf 22} (1981), 705-714.

\item{[2]} G. Kaiser,  Quantized Fields in Complex Spacetime, Ann.
Phys. {\bf 173} (1987) 338-354.

\item{[3]} G. Kaiser, {\sl Quantum Physics, Relativity, and Complex
Spacetime: Towards a New Synthesis,\/} North--Holland, Amsterdam, 1990.

\item{[4]} G. Kaiser, Generalized Wavelet Transforms.  Part I: The
Windowed X--Ray Transform, Technical Reports Series \#18, Univ. of
Lowell, 1990;  Part  II: The Multivariate Analytic--Signal  Transform,
Technical Reports Series \#19, Univ. of Lowell, 1990.

\item{[5]} G. Kaiser and R. F. Streater, Windowed Radon Transforms,
Analytic Signals and the Wave Equation, in {\sl Wavelets---A Tutorial in
Theory and Applications,\/}  C.~K.~Chui, ed. Academic Press, New York, 1992.

\item{[6]}  G. Kaiser, A Wavelet Formulation of Electrodynamics, in
preparation.

\item{[7]} L. Gross,   Norm Invariance of Mass--Zero Equations under the
Conformal Group, J. Math. Phys. {\bf 5} (1964) 687-695.

\item{[8]} H. Bateman, Proc. London Math. Soc. {\bf 8} (1910) 223, 469;
E. Cunningham, {\sl ibid.,\/} (1910) 77.

\item{[9]} R. F. Streater. and A. S. Wightman, {\sl PCT, Spin and
Statistics, And All That,\/} Benjamin, New York, 1964.

\item{[10]} I. Daubechies, A. Grossmann and Y. Meyer,  Painless
nonorthogonal expansions,  J. Math. Phys. {\bf 27} (1986), 1271-1283.

\item{[11]} I. Daubechies, {\sl Ten Lectures on Wavelets, \/} SIAM, to
appear.

\item{[12]}  G. Kaiser, An Algebraic Theory of Wavelets.  Part I:  Operational
Calculus and Complex Structure, SIAM J. Math. Anal. {\bf 23} (1992),
222-243.

\item{[13]} S. Mallat, Multiresolution Approximation and Wavelets, Trans.
Am. Math. Soc.  {\bf 315} (1989) 69-88.

\item{[14]} Y. Meyer, {\sl Ondelettes, fonctions splines et analyses
gradu\'ees,\/} lectures given at the Univ. of Torino, Italy (1986). 

\item{[15]} E. W. Aslaksen and J. R. Klauder,  Unitary representations of
the affine group,  J. Math. Phys. {\bf 9 } (1968), 206-211;
 Continuous representation  theory
using the affine group,  J. Math. Phys. {\bf 10 } (1969), 2267-2275.

\item{[16]}  T.  Paul,  Functions analytic on the half--plane as quantum
mechanical states,  J. Math. Phys. {\bf 25 } (1984), 3252-3263.

\bye